\documentstyle[a4,12pt]{article}

\newcommand{\fc}{\frac{i}{\hbar}}
\newcommand{\pg}{$P^\infty G$}

\begin{document}

\begin{center}

{\LARGE Geometric Classification of Topological\\[0.2cm] Quantum Phases}
\vspace{1cm}

{\large\sc Christopher Kohler}\footnote{E-mail: {\tt ph11ahck@rz.uni-sb.de}}
\vspace{1cm}

{\small\it Fachrichtung Theoretische Physik, Universit\"at des Saarlandes,}\\
{\small\it Postfach 151150, 66041 Saarbr\"ucken, Germany}

\end{center}

\begin{abstract}
On the basis of the principle that topological quantum phases arise from the scattering 
around space-time defects in higher dimensional unification, a geometric model is 
presented that associates with each quantum phase an element of a transformation group.  
\end{abstract}


\section{Introduction}
Within a physical theory there are often effects or phenomena studied in idealized form in 
order to gain insight into the peculiarities of the theory. In Quantum Mechanics one such 
effect is the appearance of topological quantum phases in the wave functions of particles 
moving freely in multiply connected space-times the prototype of this effect being the 
Aharonov-Bohm (AB) effect \cite{aha}, the appearance of a phase factor in the wave function 
of an electron which moves around a magnetic flux line. A similar effect is the 
Aharonov-Casher (AC) effect \cite{aha2} which is obtained from the AB effect by replacing 
the flux line and the electron by a charged line and a neutral particle with magnetic 
moment, respectively. There have also been studied analogous effects in gravitation 
\cite{dow}\cite{ana}\cite{law}\cite{ford}\cite{rez} and in non-Abelian gauge theory 
\cite{wu}. Moreover, there have been considered quantum phases associated with higher 
multipole moments of charges \cite{chen}.

It seems that topological quantum phases appear generically in theories that allow a 
geometric, gauge theoretic formulation. Up to now, however, a unified description is 
lacking.

In this letter, a combined formulation of topological quantum phases by means of a general 
model is proposed. This model provides a classification and --- to a certain extent --- 
also a prediction of topological quantum phases. The basic idea of this letter is the view 
of the ''flux line'' that generates the multiple connectedness of space-time as a 
topological defect similar to a crystal line defect. We use defects in higher dimensional 
space-times in order to describe internal gauge interactions in the framework of higher 
dimensional unification. The model is formulated as a gauge theory generalizing gauge 
theory models of gravitation in that curvatures of higher order are introduced.

The letter is organized as follows: In section 2, we begin --- as motivation and 
illustration of the model --- with a comparison of AB and AC effects in  electromagnetism 
and gravitation. In section 3, we formulate the model in a general framework and give 
some examples. Section 4 contains a summary and some comments.


\section{AB and AC Effect in Electromagnetism and Gra\-vitation}
In the electromagnetic AB effect, the wave function of a charge $q$ experiences a phase 
change when the charge moves around a magnetic flux line. The phase factor is given by
\begin{equation}\label{1}
     \Lambda^{AB}_{em} = \exp\left( \fc\oint qA_\mu dx^\mu\right)
                       = \exp\left( \fc q\phi \right), 
\end{equation}
where $A_\mu$ ($\mu =0,\ldots ,3$) is the 4-vector potential of the flux line with flux 
$\phi$ and the integration is along an arbitrary curve surrounding the flux line. The 
field strength $F_{\mu\nu}=2\partial_{[\mu}A_{\nu]}$ is singular on the flux line.

The gravitational AB effect arises when the wave function of a massive particle encircles 
a spinning cosmic string. Such a string is conveniently described by a singularity of 
torsion \cite{let}. The phase factor is
\begin{equation}\label{2}
     \Lambda^{AB}_{gr} = \exp\left( \fc\oint p_a e^a_\mu dx^\mu\right) 
                       = \exp\left( \fc 8\pi GmS \right).
\end{equation}
Here, $p_a$ ($a=0,\ldots ,3$) is the momentum of the particle with mass $m$, $e^a_\mu$ 
represents the vierbein of the string geometry with spin $S$ per unit length, and $G$ is 
the gravitational constant. $e^a_\mu$ plays the role of a potential for the torsion 
$T^a_{\mu\nu} = 2\partial_{[\mu} e^a_{\nu]}$. Since there is no curvature present, we can 
use a teleparallel formulation of gravitation and choose a gauge in which the Lorentz 
connection vanishes identically.

Comparing the two phases (\ref{1}) and (\ref{2}) we see that they have a similar form. 
Indeed, we can combine these phases into a single phase if we use a unification of 
gravitation and electromagnetism through a 5-dimensional teleparallel gravitation akin to 
the Kaluza-Klein model \cite{lee}. However, we do not employ a 5-dimensional metric. On 
the manifold $M_4\times S^1$ where $M_4$ is space-time we introduce a f\"unfbein $E^A_M 
(A,M=0,\ldots ,3,5)$ with components
\[
     E^a_\mu = e^a_\mu,\quad E^a_5=0,\quad E^5_\mu=A_\mu,\quad E^5_5=1 .
\]
In this case, the 5-th component $T^5_{\mu\nu}$ of the torsion tensor is the field 
strength $F_{\mu\nu}$. Since the charge $q$ represents the 5-th component of the 
5-momentum $p_A$, the unified AB phase is $\hbar^{-1}\oint\left( p_AE^A_\mu dx^\mu\right)$.

We now turn to the AC effect. Instead of the usual electromagnetic AC effect we consider 
its dual effect, that is, the scattering of an electric dipole moment from a straight line 
of magnetic monopoles the dipole being polarized along the line \cite{wil}. The phase 
factor reads
\begin{equation}\label{3}
     \Lambda^{AC}_{em} = \exp\left(\fc\oint\left( {\bf B}\times {\bf d}\right)\cdot 
                              d{\bf r}\right)
                       = \exp\left(\fc d_z\lambda\right),
\end{equation}
where ${\bf d}$ is the dipole moment, $d_z$ its $z$-component, and ${\bf B}$ the radial 
magnetic field of the monopole line which lies on the $z$-axis with magnetic charge 
$\lambda$ per unit length.

The counterpart to this effect in gravitation consists in the scattering of a spinning 
particle from a massive cosmic string with the spin polarized along the string. The phase 
factor is
\begin{equation}\label{4}
     \Lambda^{AC}_{gr} = \exp\left( \frac{i}{2\hbar}\oint J_{ab}\omega^{ab}_\mu dx^\mu
                               \right) 
                       = \exp\left( \fc 8\pi Gs_z M\right) .
\end{equation}
Here, $J_{ab}$ is the spin of the particle, $s_z$ its $z$-component, and $\omega^{ab}_\mu$ 
the Lorentz connection of the string with mass $M$ per unit length on the $z$-axis.

While these two AC effects are physically analogous their mathematical formulations are 
fundamentally different: The gravitational phase factor (\ref{4}) represents the holonomy 
of a locally flat Lorentz connection. The electromagnetic phase factor (\ref{3}), however, 
cannot be viewed as the holonomy of a locally flat connection. This discrepancy is 
resolved in the following way: 

The gravitational AC effect was formulated by means of a linear connection the cosmic 
string representing a curvature singularity. We consider instead a teleparallel 
formulation of gravitation in which the curvature is set to zero but a nonvanishing 
torsion is allowed. The interference of neutral spin-$\frac{1}{2}$ particles in 
gravitational fields with torsion was investigated in \cite{ana2}. In the teleparallel 
case, the phase operator is
\begin{equation}\label{5}
     \Lambda_{gr} = {\cal P}\exp\left( -\frac{i}{2\hbar}\oint \hat{S}^{ab}e_a^\mu e_b^\nu 
                       T_{\rho\mu\nu} dx^{\rho}\right), 
\end{equation}
where $\hat{S}^{ab}$ is the spin operator, $T^\rho{}_{\mu\nu}=e^\rho_aT^a_{\mu\nu}$ the 
torsion tensor, and $e^\mu_a$ the inverse of $e^a_\mu$. Roman indices are raised and 
lowered with the Minkowski metric $\eta_{ab} =\mbox{diag}(-1,1,1,1)$ or its inverse, greek 
indices with the space-time metric defined by $g_{\mu\nu}=e^a_\mu e^b_\nu\eta_{ab}$. A 
solution for a massive straight cosmic string in teleparallel gravitation is given by the 
vierbein
\begin{equation}\label{6}
     e^0 = dt, \quad e^1 = r^{-4GM} dx, \quad e^2 = r^{-4GM} dy, \quad e^3 = dz,  
\end{equation}
and the Lorentz connection $\omega^{ab}_\mu=0$ ($r^2=x^2+y^2$).  Equation (\ref{6}) is 
equivalent to the solution of a massive particle in (2+1)-dimensional teleparallel 
gravitation \cite{kaw}. Inserting the solution (\ref{6}) into the phase operator (\ref{5}), 
we recover the phase factor (\ref{4}) of the gravitational AC effect where $\hat{S}^{ab}$ 
has the only nonvanishing eigenvalue $S^{12}=s_z$. Returning to the electromagnetic AC 
effect we can write the phase factor (\ref{3}) covariantly as
\begin{equation}\label{7}
     \Lambda^{AC}_{em} = \exp\left( \frac{i}{\hbar}\oint d^\mu F_{\mu\nu} dx^\nu\right) 
             = \exp\left( \frac{i}{\hbar}\oint d^a e_a^\mu T^5_{\mu\nu}dx^\nu\right),
\end{equation}
where we have finally written the field strength $F_{\mu\nu}$ as the 5-th component of the 
torsion tensor in the 5-dimensional unification introduced above. Moreover, we have 
referred the dipole moment to the vierbein $e^a_\mu$. The formal similarity of expression 
(\ref{7}) with expression (\ref{5}) --- together with the interpretation of the field 
strength as torsion in a 5-dimensional unification --- suggests that Maxwell's formulation 
of electromagnetism represents a teleparallel formulation provided the view of a higher 
dimensional unification is adopted. This is the reason why the electromagnetic ACeffect is 
formulated differently from the gravitational one. Since the gravitational AC effect 
admits a formulation involving only curvature the same should be possible for the 
electromagnetic effect. This is indeed the case as the following considerations show:

Both a massive and a spinning straight cosmic string represent space-time defects 
\cite{gal}. These defects can be thought of as being created through global cutting and 
pasting processes (Volterra process) in which space-time points are identified by means of 
symmetry transformations. The geometries of defects can be described by locally flat 
connections associated with the groups of symmetry transformations.

A spinning cosmic string along the spatial $z$-axis is a space-time defect in the 
($z,t$)-plane. It can be generated by cutting space-time along a hypersurface bounded by 
the ($z,t$)-plane and identifying the borders after a mutual translation in time direction. 
In the terminology of the theory of crystal defects, this defect is a screw dislocation 
with Burgers vector in time direction. 

From a 5-dimensional point of view, also a magnetic flux line can be considered as a 
topological defect. In this case, 5-dimensional space-time is cut along a hypersurface 
bounded by the flux line and the cut surfaces are identified after a constant mutual 
$U(1)$-transformation of the internal $S^1$-space has been performed. If this 
transformation is viewed as a translation, the flux line corresponds to a screw 
dislocation.

A massive straight cosmic string has its counterpart in crystal physics in a wedge 
disclination. Its geometryis generated by identifying points related by a rotation around 
the string. Equivalently, it can be thought of as being created through removal of a wedge 
from space. The geometry is described in a natural way by a linear connection which has a 
curvature singularity on the string. In the teleparallel formulation of the massive
cosmic string, the defect generating rotation is considered as being local translations 
spread out over space. The resulting geometry can be illustrated by a continuous 
distribution of dislocations parallel to the string these being, however, edge dislocations 
which are created in the Volterra process through translations perpendicular to the 
defect line.

In the comparison of the gravitational AC effect with the electromagnetic one, we have 
seen above that both effects have a similar description in a teleparallel formulation. 
This suggests that also a line of magnetic charges represents a topological defect, being 
associated --- like a wedge disclination --- with a linear transformation. The magnetic 
field of a line of monopoles corresponds to a continuous distribution of radially outgoing 
flux lines which we have interpreted as screw dislocations. In the same way as the 
rotation that generates a wedge disclination can be considered as local translations, the 
local $U(1)$-transformations that generate the magnetic field of the monopole line can be 
regarded as a single linear transformation. In fact, this transformation is an internal 
$U(1)$-transformation linear in the $z$-coordinate if the monopole line is directed along 
the $z$-axis. A linear connection which describes a line of monopoles as a curvature 
singularity is associated to this linear transformation. The electromagnetic AC phase can 
be looked upon as the holonomy of this connection.

It should be remarked that also a line of electric charges can be interpreted as a 
disclination if the 4-vector potential of the dual field strength is used as will be 
described below.

To summarize this section, we have shown that the AB and AC effects in electromagnetism 
and gravitation can be regarded as being associated with topological space-time defects. 
In the following section we will generalize this result. 


\section{Classification Model}
Motivated by the considerations in the previous section we will formulate in this section 
a mathe\-matical model for the description of topological quantum phases based on the 
principle that the ''flux lines'' in the effects represent topological defects.

To this end we consider defects on a ($4+D$)-dimensional manifold $M_4\times G$ where 
$G$ is a $D$-dimensional Lie group which defines the internal interaction. A defect on 
this manifold will be thought of as being generated in a generalized Volterra process in 
the following way: The manifold ${\cal M}\times G$ with the Minkowski space ${\cal M}$ is 
cut along a hypersurface. One of the cut faces is displaced by a transformation of 
${\cal M}\times G$ and the hypersurfaces obtained are identified where possibly space must 
be added or removed. The resulting defect represents the boundary of the hypersurface and 
is of dimension $4+D-2$. We limit ourselves to defect topologies that are 2-dimensional in 
space-time.

The model employs a particular transformation group of ${\cal M}\times G$ which we denote 
by \pg. This group consists of Poincar\'e transformations of ${\cal M}$ as well as 
internal $G$-transformations that are functions on ${\cal M}$. The vector fields 
generating \pg\ are given by
\begin{equation}\label{8}
     P_a = \partial_a ,\qquad J_{ab}=x_a\partial_b - x_b\partial_a,\qquad   
               S^{(k)a\cdots d}_\alpha =\underbrace{x^a\cdots x^d}_{{k-{\rm times}}} 
                    v_\alpha ,\quad k=0,1,2,\ldots \quad ,  
\end{equation}
where $x^a$ are Cartesian coordinates on ${\cal M}$ and $v_\alpha\; (\alpha=1,\ldots ,D)$ 
are the generators of $G$ (a basis of left invariant vector fields on $G$). The vector 
fields (\ref{8}) satisfy the following commutation relations:
\begin{eqnarray}\label{9}
     [J_{ab},J_{cd}]=2\eta_{a[c} J_{d]b}-2\eta_{b[c} J_{d]a},\qquad  
     [J_{ab},P_c]=2\eta_{c[b} P_{a]},\qquad   
     [P_a,P_b] = 0,   \\[0.4cm]
                \label{10}
     [J_{ab},S^{(k)cd\cdots f}_\alpha]=2k\,\delta^{(c}_{[a}\, S^{(k)\;\;d\cdots
                      f)}_{\;\;\;\;b]\alpha} , \qquad
     [P_a,S^{(k)bc\cdots f}_\alpha]=k\,\delta^{(b}_a\,S^{(k-1)c\cdots f)}_\alpha ,\\[0.4cm] 
                \label{11}
     [S^{(k)a\cdots c}_\alpha,S^{(l)d\cdots f}_\beta]=c^\gamma_{\alpha\beta}\, 
                      S^{(k+l)a\cdots cd\cdots f}_\gamma ,\quad
                        k,l=0,1,2,\ldots\quad .
\end{eqnarray}
Here, $c^\gamma_{\alpha\beta}$ are the structure constants of $G$ and round brackets 
denote symmetrization. The first three commutators form the Poincar\'e algebra. In the 
special case that $G$ is Abelian, the commutators (\ref{11}) vanish and we can define the 
finite dimensional group $P^nG$ which is generated by the generators (\ref{8}) where 
$S^{(k)a\cdots d}_\alpha=0$ for $k>n$. If we omit the generators $P_a$ and $S^{(0)}_\alpha$ 
from (\ref{8}), the remaining vector fields generate the subgroup $P^\infty_0G$ of \pg, or 
the subgroup $P^n_0G$ of $P^nG$ if $G$ is Abelian. Since the generators $S^{(0)}_\alpha$ 
are constant on ${\cal M}$, we treat them on the same footing as the translations $P_a$.

Our aim is to describe defects within the framework of differential geometry. If on a 
manifold there is given a globally flat connection $\Gamma_0$ with respect to a 
transformation group $H$ as structure group, the Volterra process gives rise to a locally 
flat connection $\Gamma$ as long as the transformation in the Volterra process is in $H$. 
The defect geometry is characterized by nontrivial holonomies of $\Gamma$. In the case at 
hand, we therefore seek locally flat connections with the group \pg\ as structure group. 
We will follow the procedure of gauge theories of gravitation in that we consider Cartan 
connections \cite{hehl}. These arise from \pg\ connections through a symmetry breaking 
\pg\ $\to P^\infty_0G$ and have the advantage that the translational part can be related 
to a basis of cotangent space. Assume that on a principle fibre bundle $P$ over $M_4\times 
G$ with structure group $P^\infty_0G$ a Cartan connection with connection form $\omega$ 
taking values in the Lie algebra of \pg\ is given. By means of a section $s$ of
$P$ we can define a gauge connection 1-form $A=s^*\omega$ on $M_4\times G$ as the pull-back 
of $\omega$. We can decompose $A$ in the following way:
\begin{equation}\label{12}
     A=e^aP_a +\sigma^{(0)\alpha}S^{(0)}_\alpha +\frac{1}{2}\omega^{ab}J_{ab} +
                \sigma^{(1)\alpha}_aS^{(1)a}_\alpha +
                 \sigma^{(2)\alpha}_{ab}S^{(2)ab}_\alpha +\cdots\quad ,
\end{equation}
where the 1-forms $e^a$ and $\sigma^{(0)\alpha}$ are a basis of the cotangent space at 
each point of $M_4\times G$. The field strength $F=dA + A\wedge A$ is written as
\begin{equation}\label{13}
     F=T^aP_a+K^{(0)\alpha}S^{(0)}_\alpha +\frac{1}{2}R^{ab}J_{ab}+K^{(1)\alpha}_a
         S^{(1)a}_\alpha +K^{(2)\alpha}_{ab}S^{(2)ab}_\alpha +\cdots .
\end{equation}
Here, $T^a$ and $ K^{(0)\alpha}$ are torsion tensors, $R^{ab}$ and $ K^{(1)\alpha}_a$ 
curvature tensors, and $K^{(k)\alpha}_{a\cdots d}$ for $k>1$ will be referred to as 
curvature tensors of higher order.

Locally flat defect connections are characterized by $F=0$ on the manifold $M_4\times 
G\setminus\Sigma$ where $\Sigma$ is the subspace of the defect. These field equations read 
in components:
\begin{equation}\label{14}
     T^a=de^a +\omega^a{}_b\wedge e^b =0,\qquad R^{ab} =d\omega^{ab} +
                      \omega^a{}_c\wedge\omega^{cb} =0 ,  
\end{equation}
\vspace{-0.7cm}
\begin{eqnarray}\nonumber
      K^{(k)\alpha}_{ab\cdots de\cdots g} & = & d\sigma^{(k)\alpha}_{a\cdots g}
       +\frac{1}{2} c^\alpha_{\beta\gamma}\sum_{l=0}^k\sigma^{(l)\beta}_{(a\cdots d}\wedge
           \sigma^{(k-l)\gamma}_{e\cdots g)}\\[0.4cm] 
                \label{15}
      & & {}- k\;\omega^h{}_{(a}\wedge\sigma^{(k)\alpha}_{b\cdots g)h} -
         (k+1)\;\sigma^{(k+1)\alpha}_{a\cdots gh}\wedge e^h =0,\quad k=0,1,2,\ldots\quad .
\end{eqnarray}

Given a solution to these equations, we can compute the holonomy
\[
     \Lambda (*,C)={\cal P}\exp\left( -\oint_C A\right)
\]
along a closed curve $C$ in $M_4\times G \setminus\Sigma$ with base point $*$. $\Lambda$ 
is invariant under deformations of $C$ as long as the base point is held fixed.

We are now in a position to formulate the model for the classification of topological 
quantum phases: Given a 1-parameter subgroup of the group \pg\ generated by the vector 
field $v$ on $M_4\times G$, we associate to it a quantum mechanical operator $\hat{v}=i
\hbar v$ which is interpreted as the charge operator of the quantum mechanical system that 
encircles the ''flux line'' in the interference experiment. In the Volterra process, the 
1-parameter subgroup generates a defect the gauge field of which can be determined with 
the help of the field equations (\ref{14},\ref{15}) where $F$ has a $v$-valued singularity 
concentrated on $\Sigma$. The holonomy of the gauge connection is interpreted as a phase 
operator acting on the wave function of the quantum mechanical system. We require that 
the wave function is an eigenfunction of the operator $\hat{v}$. The phase operator then 
becomes a phase factor which gives the topological quantum phase.

With each of the generators (\ref{8}) there is associated a charge. The hierarchy of the 
generators $S_\alpha^{(k)a\ldots d}$ corresponds to the hierarchy of multipole moments of 
the charge $S_\alpha^{(0)}$. Given a topological quantum phase, we can find the 
1-parameter group that characterizes the quantum mechanical system that interferes as well 
as the ''flux line'' which represents a topological defect where the group parameter gives 
its strength. On the other hand, choosing a 1-parameter subgroup of \pg\ with a given $G$ 
a new quantum phase can be determined. In this case, it is, however, not ensured that 
this phase is realized in nature since the model is purely topological and does not take 
into account the real interactions. For example, whether the quantum mechanical systems 
experience classical forces cannot be predicted from the model.

We will give a few examples to the model:

(1) Let $G$ be the trivial group $I$. In this case, $P^\infty I$ is the Poincar\'e group 
and the field equations reduce to the equations (\ref{14}). The defects that can be 
generated by means of Poincar\'e transformations in the Volterra process are space-time 
dislocations and disclinations as explained in section 2. The associated charges are mass 
and spin, leading to the gravitational AB and AC effect, respectively.

(2) We consider $G=U(1)\times U(1)$ corresponding to electromagnetism with magnetic and 
electric flux. We limit ourselves to the group $P^2(U(1)\times U(1))$ and require further 
that the Lorentz connection is flat and torsionfree choosing $\omega^a{}_b =0, e^a=dx^a$. 
The field equations (\ref{14},\ref{15}) then reduce to
\begin{eqnarray}\nonumber
     K^{(0)\alpha} =d\sigma^{(0)\alpha} -\sigma^{(1)\alpha}_a\wedge dx^a =0,\\[0.5cm]
                \nonumber
     K^{(1)\alpha}_a =d\sigma^{(1)\alpha}_a -2\sigma^{(2)\alpha}_{ab}\wedge dx^b =0,
                \\[0.5cm]\label{20}
     K^{(2)\alpha}_{ab} =d\sigma^{(2)\alpha}_{ab} =0,\qquad \alpha =1,2.
\end{eqnarray}

(2a) In the case of the AB effect the quantum mechanical system is an electrically or 
magnetically charged particle. Thus, the associated 1-parameter subgroup of $P^2(U(1)
\times U(1))$ consists of internal $U(1)$ transformations constant on $\cal M$. Only the 
torsion $K^{(0)\alpha}$ is nonvanishing and singular on the flux line where $\alpha=1$ 
corresponds to magnetic flux and $\alpha=2$ to electric flux. These flux lines are screw 
dislocations on the space ${\cal M}\times S^1\times S^1$. Choosing $\sigma^{(1)\alpha}_a 
=\sigma^{(2)\alpha}_{ab}=0$ the holonomy of the connection $\sigma^{(0)\alpha}$, which 
represents the magnetic ($\alpha=1$) or electric ($\alpha=2$) 4-vector potential, gives 
the AB phase factor.

(2b) For the AC effect the interfering system is a dipole moment. The associated 
1-parameter subgroups of $P^2(U(1)\times U(1))$ are $U(1)$ transformations which are 
linear on $\cal M$. The corresponding defects are disclinations characterized by 
singularities of the curvature $K^{(1)\alpha}_a$ with $\alpha = 1$ for electric dipoles 
and $\alpha = 2$ for magnetic ones. The field equations (\ref{20}) are solved by
\begin{equation}\label{21}
     \sigma^{(2)\alpha}_{ab} =0,\qquad \sigma^{(1)\alpha}_a =\frac{k^\alpha_a}{2\pi} 
           d\varphi ,\qquad\sigma^{(0)\alpha} =d\theta^\alpha -\frac{k^\alpha_a}{2\pi} 
              x^a d\varphi , 
\end{equation}
where the defect lies along the $z$-axis and $\varphi$ is the azimuthal angle. $k^\alpha_a$ 
is the defect (or group) parameter and $\theta^\alpha$ are (angular) coordinates on 
$S^1\times S^1$. In the case that only $k^\alpha_3$ is nonvanishing, the holonomy of the 
connection (\ref{21}) gives the AC phase factor. Alternatively, we can use a teleparallel 
formulation setting $K^{(1)\alpha}_a = 0$. Then, with $\sigma^{(1)\alpha}_a = 0$, 
$\sigma^{(0)\alpha}$ in equation (\ref{21}) gives a nonvanishing torsion $K^{(0)\alpha}$ 
which is the field strength ($\alpha =1$) or the dual field strength ($\alpha =2$). If 
only $k^\alpha_3$ is nonvanishing, $K^{(0)1}$ is the field strength of a homogeneous 
magnetic line charge and $K^{(0)2}$ that of an electric one.

(2c) In the case that the curvature of second order $K^{(2)\alpha}_{ab}$ is singular on 
the defect, we obtain a topological quantum phases for quadrupole moments. The ''flux 
line'' corresponds to a defect which results in the Volterra process from a $U(1)$ 
transformation quadratic on $\cal M$. Again, a teleparallel formulation is possible where 
we have a nonvanishing $K^{(0)\alpha}$.


\section{Conclusion}
In this letter, we have proposed a model that allows a classification of topological 
quantum phases in that an element of a transformation group of a higher dimensional 
space-time is associated to each quantum phase. Our starting point was the principle that 
topological quantum phases arise in the scattering from space-time defects. The phase 
factors are then given by the holonomies of the defect geometries. The model provides 
moreover an explanation why some quantum phases in electromagnetism are usually not 
described by holonomies of locally flat connections: From the viewpoint of a higher 
dimensional unification, Maxwell's theory possesses the nature of a teleparallel theory.

We close with a few remarks:

(1) The gauge fields of higher order we have introduced do not seem to be suitable for a 
formulation of a dynamics of gauge fields in the general case. They are only used here to 
show that the "flux lines" in the topological quantum phases represent space-time defects. 
Kaluza-Klein theory gives a formulation of electromagnetism in terms of linear connections, 
the corresponding symmetries are, however, broken.

(2) A crucial property of the model is the combination of external and internal 
transformations in the gauge group. As a result, the field equations (\ref{14},\ref{15}) 
of different order are coupled. This has the consequence that a defect described by a 
curvature of a given order can be represented as a pair (dipole) of defects of one order 
higher.

(3) At least in the case that $G$ is Abelian, there is a close relation between principle 
\pg-bundles and bundles of frames of higher order over $M_4\times G$ \cite{kob}.

(4) There exists another attempt to formulate the electromagnetic AC effect by means of a 
holonomy of a connection \cite{ana3}\cite{oh} using the fact that a neutral particle with 
magnetic moment in an electromagnetic field is equivalent to an isospin particle in an 
$SU(2)$ gauge field \cite{gol}\cite{froe}. However, the $SU(2)$ connection is not flat and 
cannot be interpreted as describing a topological defect. Moreover, the $SU(2)$ symmetry 
originates from the spin of the particle.


\section*{Acknowledgement}
I would like to thank Th.Filk and H.M.Sauer for useful discussions. This work was 
supported by the Deutsche Forschungsgemeinschaft (DFG) through Grant No.\ Ho 841/9-2.


\end{document}